\documentstyle[aps,12pt]{revtex}
\begin{document}
\author{R.J. Radwa\'{n}ski}
\address{Center for Solid State Physics, \'{s}w. Filip 5, 31-150 Krak\'{o}w, Poland,
e-mail: sfradwan@cyf-kr.edu.pl}
\title{FORMATION OF THE HEAVY-FERMION STATE \\
- AN EXPLANATION \\
IN A MODEL TRADITIONALLY CALLED LOCALIZED$^{{\bf ***}}$}
\maketitle

\begin{abstract}
In contrary to widely spread view about the substantial delocalization of $f$
electrons in heavy-fermion (h-f) compounds it is argued that h-f phenomena
can be understood with localized $f$ electrons. Then the role of
crystal-field interactions is essential and the heavy-fermion behaviour can
occur for the localized Kramers-doublet ground state.
\end{abstract}

\date{16.06.1999}

In the proposed explanation compounds exhibiting heavy-fermion (h-f)
behaviour are considered - in analogy to normal rare-earth intermetallics -
within a few electronic subsystems. For the understanding of the h-f
behaviour it is essential to distinquish {\it f} electrons from conduction
electrons. These two subsystems are independent as far as the direct
electron hopping from one to other subsystem does not occur. The {\it f}
subsystem is a highly correlated {\it f}$^{\text{ }n}$ electonic system. The
proposed model takes advantage of two recent findings: i)
crystalline-electronic-field (CEF) interactions of the f shell can produce a
non-magnetic ground state even in case of the Kramers system (the analytical
proof exists, at present, for the {\it f}$^{\text{ 3}}$ system in the
hexagonal symmetry [1]) and ii) the {\it f}$^{\text{ }n}$ localized states
for an intermetallic compound containing an {\it f} atom always lie at the
specific-heat probing level (the virtual Fermi level) as the {\it f}$^{\text{
}n}$ states are many-body states in contrary to single-electron states
within the conduction-electron band. The different nature of excitations
allows for independent contributions of these two subsystems to magnetic and
electronic properties. The CEF state

\begin{center}
$\Gamma _9$=$\frac{\sqrt{3}}2$ 
%TCIMACRO{\TEXTsymbol{\vert} }
%BeginExpansion
\mbox{$\vert$}
%EndExpansion
$\pm $3/2%
%TCIMACRO{\TEXTsymbol{>} }
%BeginExpansion
\mbox{$>$}
%EndExpansion
+ $\frac 12$ 
%TCIMACRO{\TEXTsymbol{\vert}}
%BeginExpansion
\mbox{$\vert$}
%EndExpansion
$\mp $9/2%
%TCIMACRO{\TEXTsymbol{>}}
%BeginExpansion
\mbox{$>$}
%EndExpansion
\end{center}

given for the {\it f}$^{\text{ 3}}$ subsystem in the hexagonal CEF
interactions is a Non-Magnetic Kramers doublet as expectation values for J$%
_x $ J$_y$, and J$_z$, of the total angular momentum are equal zero. In ref.
1 this state has been proved to be realized as the ground state. In
compounds exhibiting the heavy-fermion behaviour, the ground sate of the 
{\it f}-electron subsystem tends to the N-M Kramers doublet. Then one has in
the single-ion picture enhanced but finite susceptibility at 0 K and a
normal Curie-Weiss behaviour at higher temperatures exactly as is
experimentally observed. The N-M Kramers doublet ground state of {\it f}
electrons behaves like a half-filled band at the Fermi level (2 states and 1 
{\it f}$^{\text{ }n}$ particle) allowing for a ''delocalization'' of the 
{\it f} electrons and, in particular, for a many low-energy excitations
detected as an enormous specific heat at lowest temperatures. In the
presented model correlations between f electrons and conduction electrons
proceed via electrostatic in interactions. The heavy-fermion state results
from competition between CEF and antiferromagnetic interactions.

Some further implications of the model will be discussed.

$^{***}$presented at International Conference on Strongly Correlated
Electron Systems, Sendai, Japan, September 7-11, 1992 as the poster 8P-92.
The above text has been printed in the abstract booklet. The paper has been
rejected from publication by the Organizing Committee despite of the strong
author's complains.

\end{document}